# The Automatic Neuroscientist:
# automated experimental design with real-time fMRI

Romy Lorenz[1, 2], Ricardo Pio Monti[3], Inês R. Violante[1], Christoforos Anagnostopoulos[3], Aldo A. Faisal[2], Giovanni Montana[3, 4] & Robert Leech[1]

[1] The Computational, Cognitive and Clinical Neuroimaging Laboratory, The Division of Brain Sciences, Imperial College London, London, UK
[2] Department of Bioengineering, Imperial College London, London, UK
[3] Department of Mathematics, Imperial College London, London, UK
[4] Department of Biomedical Engineering, King's College London, London, UK

**Abstract**

A standard approach in functional neuroimaging explores how a particular cognitive task activates a set of brain regions (one task-to-many regions mapping). Importantly though, the same neural system can be activated by inherently different tasks. To date, there is no approach available that systematically explores whether and how distinct tasks probe the same neural system (many tasks-to-region mapping). In our work, presented here we propose an alternative framework, the *Automatic Neuroscientist*, which turns the typical fMRI approach on its head. We use real-time fMRI in combination with state-of-the-art optimisation techniques to automatically design the optimal experiment to evoke a desired target brain state. Here, we present two proof-of-principle studies involving visual and auditory stimuli. The data demonstrate this closed-loop approach to be very powerful, hugely speeding up fMRI and providing an accurate estimation of the underlying relationship between stimuli and neural responses across an extensive experimental parameter space. Finally, we detail four scenarios where our approach can be applied, suggesting how it provides a novel description of how cognition and the brain interrelate.

## 1. Introduction

A central aim of functional neuroimaging is investigating how the brain is associated with a cognitive or perceptual task. However, the mapping between cognitive tasks and neural processes is typically not straightforward. Most tasks (particularly high-level cognitive tasks) do not map neatly onto a single neural system but simultaneously activate a myriad of regions. Equally, a given neural system is activated by many often quite different tasks. This means that to understand the relationship between the brain and cognition involves understanding a many (tasks) to many (regions) mapping. However, the standard approach involves acquiring whole-brain functional magnetic resonance imaging (fMRI) data to probe the neural systems associated with one or a few task conditions; this, therefore, only addresses one side of the many-to-many mapping (one task-to-many regions). It does not answer whether (and how) many different tasks activate similar systems (many tasks-to-region mapping), which is, in many ways, a much more difficult problem.

In the literature, the same network of brain regions have frequently been ascribed with completely different functional descriptions based on the different tasks that evoked them, e.g., "the pain matrix" (evoked by painful stimuli, see [1]–[3]) is very similar to the "salience network" evoked by cognitively surprising stimuli [4], [5]. Similarly, the superior temporal sulcus has been termed the "chameleon of the brain" [6] due to its involvement in different functional roles, ranging from audio-visual integration [7] to motion [8], speech [9] and face processing [10]. Therefore, from a theoretical perspective, only considering how one or a few tasks activate a given region makes understanding its functional role as difficult as looking for a needle in a haystack; we do not know which task activates a region and how it compares to all the other possible tasks. It can also give rise to the reverse inference problem [11], where cognitive functions (particularly sub-processes hypothesized to be involved in a larger cognitive process) are ascribed to a region because that region has activated for a task previously. For example, based on greater activity in rats' striatum during pup suckling versus cocaine administration, a study concluded that "pup suckling is more rewarding than cocaine" [12]. This interpretation was inferred from activation in a brain region that had previously been reported to engage in reward processing, rather than by being directly manipulated (example from [11]).

It is not just when looking at functional mappings between tasks and cognition that these problems are evident, but also when investigating individual differences in neural function or in understanding neurological or psychiatric disorders. The typical approach is to see how a given region or network or regions vary in e.g., evoked activity with individual performance or clinical dysfunction. However, an alternative way to think about differences across individuals (or groups) is to see how many tasks map onto activation of a brain region or set of regions. Individual differences in regional activation for a given task may reflect a different involvement for that region in different tasks, rather than absolutely lower or greater activity in that region.

In this work we propose and pilot an alternative framework that turns the typical fMRI approach on its head to better address the many tasks-to-region mapping: The *Automatic Neuroscientist* (see Figure 1 for an overview of our proposed method). We use real-time fMRI in combination with state-of-the-art optimisation techniques to automatically adjust the experimental conditions. This closed-loop optimisation algorithm starts with a pre-defined brain region and finds a set of tasks/stimuli that activate it. This approach is far more efficient in that it can assess many experimental dimensions simultaneously before converging on the optimal experimental setup to evoke a desired pattern of activation. The approach also efficiently develops a description of the whole parameter space (across tasks), meaning the complex relationship between task and brain can be unveiled more easily.

We have conducted two proof-of-principle studies involving visual and auditory stimuli as well as optimisation algorithms of varying complexity. The approach explores a large experimental parameter space aiming to maximize activity in visual or auditory brain regions (as desired) by changing the visual/auditory stimuli that the subjects are exposed to. The data shows this approach to be very powerful, hugely speeding up fMRI and allowing a different set of questions to be asked about how cognition and the brain are interrelated.

## 2. Methods

### 2.1. Subjects

Twelve healthy volunteers (7 females, mean age ± SD: 26.8 ± 4.5 years) participated in our study. Subjects had no history of either contraindication to MRI scanning or neurological/psychiatric disorders. All subjects had normal or corrected-to-normal vision and gave written consent for their participation. Subjects were informed about the real-time nature of the fMRI scans but no information was given on the actual aim of the study or which parameters in the experiment would be adapted in real-time. Most importantly, subjects were unaware of the target brain state our algorithm was optimising for.

### 2.2. Target brain regions

Based on a previous study [13] we identified two target brain regions: bilateral lateral occipital cortex and bilateral superior temporal cortex (see Figure 2) that were strongly activated for complex visual (e.g., naturalistic movies) or auditory stimuli (speech), respectively. Masks for these two regions of interest (ROI) were obtained from thresholded (*z* > 5) and binarized group-level maps [13].

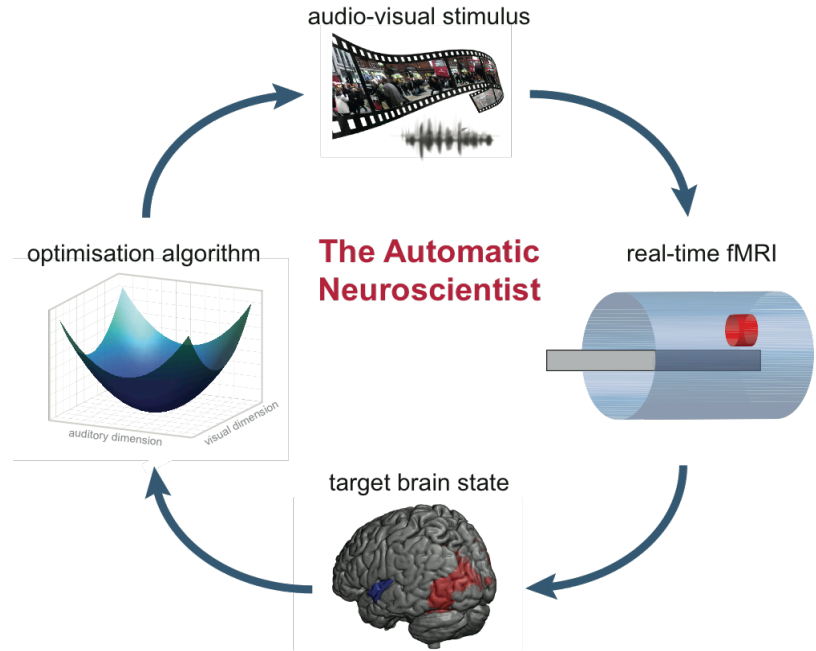

**Figure 1:** High-level overview of proposed method. The optimisation algorithm starts with selecting a random parameter vector from the available experimental parameter space. The parameter vector determines an experimental condition (e.g. an audio-visual stimulus combination) that is presented to the subject. Whole-brain functional images are acquired and pre-processed in real-time in response to the stimulus. Information from the target brain regions is extracted and then fed into the optimisation algorithm. Based on this, the optimisation algorithm chooses a parameter vector closer to the minimum of the objective function.

2.3. Real-time fMRI setup

Whole-brain coverage images were acquired in real-time by a Siemens Verio 3T scanner using an EPI sequence (T2*- weighted gradient echo, voxel size: 3.00 x 3.00 x 3.00 mm, field of view: 192 × 192 × 105 mm, flip angle: 80°, repetition time (TR) / echo time (TE): 2000/30 ms, 35 interleaved slices with 3.00 mm thickness). The reconstructed single EPI volume was exported from the MR scanner console to the real-time fMRI processing computer (Mac mini, 2.3 GHz Intel Quad Core i7, 16 GB RAM) via a shared network folder.

Prior to the online run, a high-resolution gradient-echo T1-weighted structural anatomical volume (reference anatomical image (RAI), voxel size: 1.00 × 1.00 × 1.00 mm, flip angle: 9°, TR/TE: 2300/2.98 ms, 160 ascending slices, inversion time: 900 ms) and one EPI volume (reference functional image (RFI)) were acquired.

Offline and online pre-processing were carried out with FSL [14]. The first steps occurred offline prior to the real-time fMRI scan. Those comprised brain extraction of the RAI and RFI using BET [15] followed by an affine co-registration of the RFI to the RAI and subsequent linear registration (12 degrees of freedom) to a standard brain atlas (MNI) using FLIRT [16]. The resulting transformation matrix was used to register the two target brain maps from MNI to the functional space of the respective subject.

For online runs, incoming EPI images were converted from dicom to nifti file format and real-time motion correction was carried out using MCFLIRT [17] with the previously obtained RFI acting as reference. In addition, images were spatially smoothed using a 5 mm FWHM Gaussian kernel. ROI means of the functional masks for each TR were simultaneously extracted using a general linear model (GLM) approach and written into a text file that was read by a MATLAB script carrying out a second stage of pre-processing.

The second stage of pre-processing involved cleaning the two extracted timecourses in real-time by removing low-frequency signal drifts with an exponential moving average (EMA with smoothing factor α = 0.96, time constant τ = 49 sec, high-pass filter cut-off frequency = 0.003 Hz). In addition, high-frequency noise and large signal spikes were removed with a modified Kalman filter (both algorithms were obtained from [18]). The pre-processed timecourses were written into a separate text file for subsequent analyses.

The experiment commenced after a burn-in period of 10 TRs. This allowed for more reliable pre-processed ROI timecourses (the EMA stabilized after approximately 20 seconds).

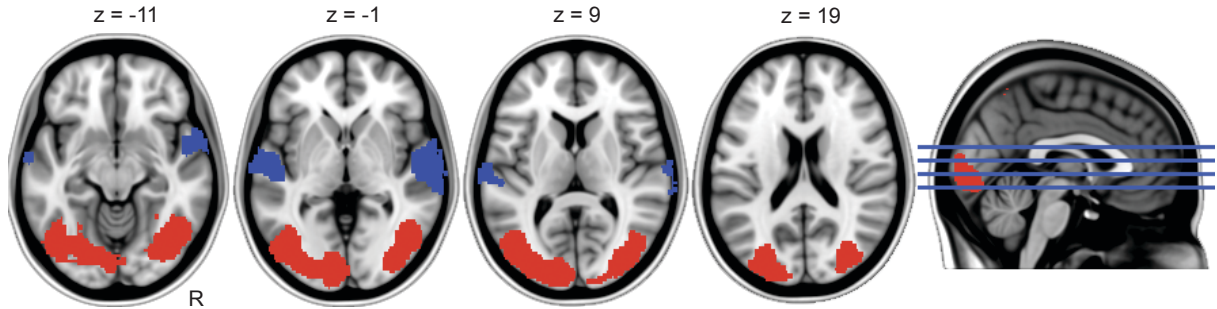

**Figure 2:** Target brain regions. The two target brain regions were the bilateral lateral occipital cortex (red) and the bilateral superior temporal cortex (blue). Masks were obtained from thresholded (*z* >5) and binarized group-level results [13].

### 2.4. Audio-visual stimuli

In both experiments, subjects were presented with audio-visual stimuli in blocks of 10 seconds followed by 10 seconds of no stimulus baseline (black background). Stimuli were presented in the centre of the screen with a black background using the MATLAB Psychophysics Toolbox (Psychtoolbox-3, [19], [20]). Subjects were instructed to actively attend to any auditory or visual stimuli without prioritizing one modality over the other. For both modalities, stimuli varied in complexity from no visual input (black background as in baseline) to a moving naturalistic street scene or from no audio input (as in baseline) to a spoken sentence by a human voice in 10 discrete steps, respectively.

**Table 1.** Overview of visual parametric stimulus variation from no (1) to complex (10) visual input

| Visual complexity | Number of frames | Image size (pixel dimensions) | Image saturation (percentage) | 2D Gaussian filter (*sd*) | Added Gaussian noise (variance) |
|---|---|---|---|---|---|
| 1 | - | - | same as baseline | same as baseline | same as baseline |
| 2 | 1 | 65 x 116 | 1.7 | 8.9 | 8.9 |
| 3 | 3 | 125 x 221 | 2.8 | 7.8 | 7.8 |
| 4 | 6 | 184 x 327 | 4.6 | 6.7 | 6.7 |
| 5 | 11 | 243 x 432 | 7.7 | 5.6 | 5.6 |
| 6 | 21 | 303 x 538 | 12.8 | 4.5 | 4.5 |
| 7 | 39 | 362 x 644 | 21.2 | 3.4 | 3.4 |
| 8 | 72 | 422 x 749 | 35.3 | 2.3 | 2.3 |
| 9 | 133 | 481 x 855 | 58.7 | 1.2 | 1.2 |
| 10 | 250 | 540 x 960 | original video | original video | original video |

*Visual stimuli.* For visual stimuli we used colour video footage displaying a naturalistic street scene previously used in [13]. Stimuli varying parametrically were generated by altering the following features of the video: number of frames (in which a lower number of frames is subjectively experienced as a slower video), video image size, image saturation, spatial blurring (using a 2D Gaussian filter of constant size with varying standard deviations *sd*) and varying amounts of added Gaussian white noise (with zero mean and varied variance) (Table 1).

*Auditory stimuli*. For auditory stimuli, we used four sentences spoken by one male speaker retrieved from the Australian National Database of spoken language [21]. Parametrically varying stimuli were generated using noise-vocoded speech, created with Praat [22]. Noise-vocoded speech was created by dividing the spoken sentence into varying numbers of logarithmically-spaced frequency bands. For each frequency band the amplitude envelope was extracted, and then used to modulate noise in the respective frequency band. Finally, the frequency bands are recombined to create the noise-vocoded sentence. To increase parametric variation amongst the stimuli, we added additional Gaussian noise with zero mean and varied the standard deviation (*sd*) to the original sentence before noise-vocoding was performed (Table 2). To reduce adaptation effects of the stimuli, we randomly selected two out of four sentences (read by the same male speaker) for each run.

**Table 2.** Overview of auditory parametric stimulus variation from none (1) to complex (10) auditory input

| Auditory complexity | Number of bands for noise-vocoded speech | Added Gaussian noise (*sd*) |
|---|---|---|
| 1 | same as baseline | same as baseline |
| 2 | 1 | .08 |
| 3 | 2 | .05 |
| 4 | 3 | .03 |
| 5 | 4 | .01 |
| 6 | 5 | .008 |
| 7 | 6 | .005 |
| 8 | 10 | .003 |
| 9 | 20 | .001 |
| 10 | original sentence | original sentence |

### 3. Study 1

For our first study, we defined a two-dimensional experiment parameter space with 100 possible states: 10 discrete steps along the visual dimensions and 10 discrete steps along the auditory dimension. For both dimensions, stimuli increased in complexity as described above (see Figure 3a). Each audio-visual stimulus presented to the subjects was chosen from this large two-dimensional parameter space by the optimisation algorithm.

#### 3.1. Scanning conditions

For the first study, seven subjects (5 females, mean age ± SD: 26.7 ± 5.3 years) underwent two separate adaptive real-time fMRI runs. The runs only differed in the a-priori defined target brain state. The two tested target brain states of interest were: (1) maximized occipital cortex activity with minimum superior temporal cortex activity; and (2) maximized superior temporal cortex with minimized occipital cortex activity. The order of runs was counterbalanced across participants. Based on prior literature [13], we strongly hypothesized that our target brain state (1) would be evoked by the most complex visual stimulus in combination with no auditory input (see red square in the bottom right corner in Figure 3a) while the reverse (complex auditory, no visual) was expected for our target brain state (2) (see blue square in the top left corner in Figure 3a).

#### 3.2. Optimisation using stochastic approximation algorithms

The objective for our first study was to traverse the two-dimensional parameter space for the algorithm to automatically learn the combination of audio-visual stimuli that best evokes a certain target brain state. Therefore, the problem can be seen as a two-dimensional optimisation problem.

There are two fundamental challenges posed here. First, the objective function (in this case the difference of blood-oxygen-level dependent (BOLD) activation between the two ROIs) is not available analytically. As a result, gradient-driven algorithms are not suitable here and *gradient-approximation* methods must be employed. Moreover, the presence of non-neural noise (with both physiological and non-physiological origin) is well documented for fMRI experiments. As a result, measurement of the objective function (i.e., estimated BOLD signal) is expected to be corrupted by noise.

In order to address both of these issues, a Simultaneous Perturbation Stochastic Approximations (SPSA, [23]) algorithm was employed. A series of modifications were made in order to address the discrete nature of our parameter space [24].

SPSA algorithms are particularly well suited for the complex nature of the experiment at hand. First, SPSA algorithms are stochastic approximations that do not require analytic information about the nature of the objective function. At each iteration, the algorithm randomly proposes two new potential selections within the experimental parameter space where the objective (brain activation, see below) is evaluated. These results are then used to obtain an approximation to the gradient, from which the algorithm proposes two selections. Video 1 demonstrates an exemplary search of the SPSA algorithms through the two-dimensional experiment parameter space.

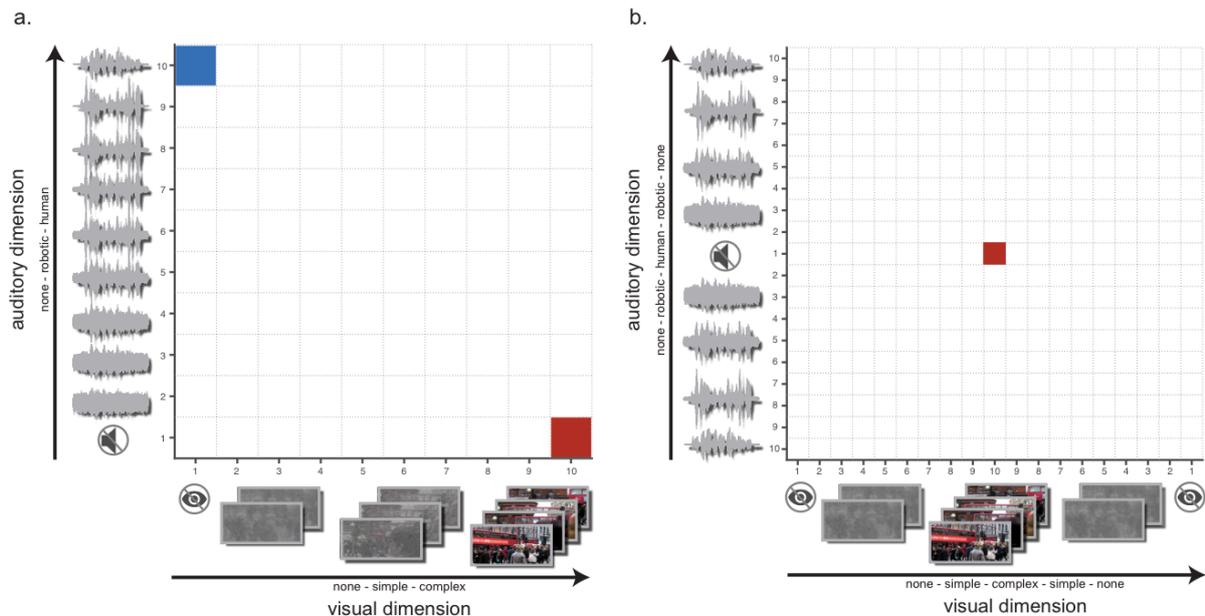

**Figure 3:** Two-dimensional experiment parameter space for both studies. **a.** Parameter space of study 1 with 10x10 (100) possible combinations composed of auditory and visual stimuli of varying complexity. The hypothesized optimal stimulus combination for evoking target brain state (1) is the most complex visual stimulus in combination with no auditory input (coloured in red). The reverse stimulus combination (complex auditory, no visual input) was hypothesized to be optimal for target brain state (2) (coloured in blue). **b.** Parameter pace of study 2 with 19x19 (361) possible combinations. The hypothesized optimal stimulus combination for evoking the target brain state (1) is the most complex visual stimulus in combination with no auditory input (coloured in red).

### 3.2.1. Objective function

After the presentation of two successive audio-visual stimuli, we estimated the BOLD signal associated with each and compared it to an a-priori defined target brain state: this difference was termed the *loss*. The loss was designed to show whether there was greater activity in the superior temporal or occipital lobe mask for the first or the second combination of audio-visual stimuli. To estimate the loss, we ran separate general linear models (GLM) on the previous 20 time points (10 TRs covering stimulus presentation and subsequent baseline presentation) of the cleaned time courses extracted from the occipital and temporal cortical masks, respectively. Each GLM consisted of an intercept term and two stimulus regressors, which were modelled by convolving a boxcar kernel with a canonical double gamma hemodynamic response function (HRF). This resulted in two estimated regression coefficients per GLM (disregarding the intercept term), representing the first and second stimulus, for which a corresponding post-hoc contrast was calculated comparing the two regression coefficients. A positive t-value indicated that the first stimulus presented more closely matched the target brain state while a negative t-value indicated that the second stimulus more closely corresponded to the target brain state. Separate t-values were calculated for the occipital and temporal masks. If the two t-values from the different masks were in conflict, the larger t-value was used to determine which of the two states was chosen to govern the next choice of the algorithm. For half the runs, the loss function was based on maximizing occipital activity and minimizing superior temporal, and for the other half this was reversed.

### 3.2.2. Stopping criterion

The SPSA algorithm is expected to eventually converge to a local optima as the number of iterations increases [23]. In this work, convergence was based on an arbitrary threshold and defined as sampling the same combination of stimuli for three consecutive iterations (i.e., the same stimuli combination was chosen to be optimal on three consecutive iterations). If convergence occurred the scan was stopped. If convergence did not occur within 10 minutes, the experiment ended automatically in order to keep scanning time to a confortable length for the participants.

### 3.3. Results

The results of all seven subjects are listed in Table 3. In total, we achieved correct convergence in 11 out of 14 runs. For one run (sub_03), the SPSA faulty converged slightly off the optimum (at [4 10] instead of [1 10]) while for two other runs, convergence did not occur within 10 minutes. In one of the runs where convergence did not occur (sub_04), the algorithm remained in the vicinity of the hypothesized optima, which was visited more frequently than any other stimulus combination. As a result, this failure to converge may be a result of substantial noise in the subjects' BOLD response.

We performed non-parametric permutation testing in order to assess how likely the pattern of results we observed could have occurred by chance. We created a null-distribution by simulating (10,000 permutations): the (a) mean rate and (b) median rate of convergence for 14 experiments (two runs per subject) when randomly assigning our empirically obtained objective function values (t-values) for each iteration within an experiment. For practical reasons, the maximal rate of convergence was set to 50 – so an experiment that did not converge within 50 iterations was set to the maximal rate. After we obtained a distribution of test values for (a) the mean rate and (b) median rate of convergence expected under the null hypothesis, the p-value associated with our observed test statistic was computed.

In order to obtain our observed test statistic, we set all runs that did not converge or faulty converged to 50 (three runs in total). This resulted in the empirical mean rate of convergence of 16.86 and a median rate of convergence of 9. The corresponding z-values for the (a) mean rate of convergence were -2.4035 ($p = 0.0061$, one-tailed) and for the (b) median rate of convergence -2.1152 ($p = 0.0134$, one-tailed). Note, that we used a very conservative approach, since in reality we stopped the experiment after 14 iterations (rather than 50). A more liberal approach for obtaining our test statistic, for example by setting all runs that did not or faulty converged to, e.g., the mean rate of convergence instead (27.4), we would have obtained an even larger z-value of -3.50 ($p = 0.00016$, one-tailed).

3.1. Limitations

This study was designed to establish the feasibility of the general approach to efficient, automatic experiment design with near real-time fMRI. This was demonstrated by the fact that the audio-visual stimulus combination rapidly converged to the hypothesized optima, much faster than if an exhaustive search were performed. However, there remains much room for improvement, in particular with respect to the use of the SPSA algorithm. Briefly, SPSA algorithms make very limited assumptions regarding the nature of the data. While this may be beneficial in many scenarios, is also comes at a cost. In particular, the use of an SPSA algorithm relegates the objectives of our approach to only learning the stimulus pairing associated with (possibly local) maxima as opposed to obtaining an understanding of the global relationships between stimuli and neural response across experiment parameters. As a result, with SPSA we are only able to obtain a very limited understanding of the relationship between the stimuli and the response. Moreover, the SPSA algorithm effectively estimates the gradient at every iteration. This increases the susceptibility of the SPSA algorithm to noisy outliers and reduces the efficiency of such an approach in low signal-to-noise scenarios.

**Table 3.** Results of all subjects from study 1 for both runs

| Subject | Target brain state (1) (iterations/minutes) | Target brain state (2) (iterations/minutes) |
|---|---|---|
| sub_01 | 10 / 6.67 min | 9 / 6 min |
| sub_02 | 13 / 8.67 min | 8 / 5.33 min |
| sub_03 | 7 / 4.67 min | *faulty convergence* |
| sub_04 | 9 / 6 min | *no convergence* |
| sub_05 | 6 / 4 min | 10 / 6.67 min |
| sub_06 | 6 / 4 min | 4 / 2.67 min |
| sub_07 | *no convergence* | 4 / 2.67 min |

## 4. Study 2

Building on Study 1, for Study 2 we modified the approach by employing a Bayesian optimisation approach to better estimate the whole parameter space and to better account for the noisy data. In the second study, we again used a similar two-dimensional experiment parameter space; however, this time we expanded the parameter space described in Study 1 by mirroring the visual and auditory axes. This resulted in a more challenging experimental parameter space with 361 possible states: 19 discrete steps along the visual dimensions and 19 discrete steps along the auditory dimension (Figure 3b).

Study 1 demonstrated that our approach works for either target brain state (maximizing occipital or temporal cortex activity). Therefore, for practical reasons and to make the design simple, Study 2 was limited to exclusively explore target brain state (1) (i.e., maximized occipital cortex activity while superior temporal cortex activity is suppressed). The hypothesized optimum (complex visual stimulus with no auditory input) was now, due to the mirroring of the axes, located in the centre of our experiment parameter space (see red square in Figure 3b).

4.1. Scanning conditions

For our second study, five new subjects were recruited (2 female, mean age ± SD 26.8 ± 3.6 years). Four of them underwent four separate real-time fMRI runs while we were only able to record a single run in one subject due to technical failure of the scanner. All runs aimed to evoke target brain state (1).

### 4.2. Real-time experimental design using Bayesian Optimisation

As discussed previously, there are several challenging aspects presented in this work. First and foremost, the goal of the proposed method is to optimize an unknown objective function. By this we mean that there is no, a priori, analytical expression available relating the activation in the target regions to the visual and auditory stimuli. Moreover, while there are hypotheses in the literature regarding the activation of these regions, we cannot make formal statements regarding aspects such as the convexity of the objective function nor can we assume we have access to its derivatives.

In order to address these issues we used a Bayesian optimisation algorithm to efficiently learn the combination of audio-visual stimuli that maximises activation in the target brain state [25]. The proposed algorithm consists of an iterative scheme where subjects are presented with a stimulus and their BOLD activation is measured. This activation is subsequently provided as feedback and incorporated in real-time to learn a probabilistic model of the subject's target region activation as a function of both the auditory and visual stimuli. The model is subsequently employed to propose a new combination of auditory and visual stimuli, which is likely to increase activation in the target regions. In this manner, the proposed algorithm is able to effectively learn the distribution of the latent objective function.

The implementation of a Bayesian optimisation algorithm requires two fundamental choices. First, a non-informative prior distribution must be specified for the latent objective function. Here, a Gaussian process prior is employed due to its flexibility and tractability [26]. Formally, Gaussian processes can be seen as an extension of the multivariate Gaussian distribution to an infinite dimensional stochastic process. Due to the marginalization properties of the Gaussian distribution, Gaussian processes are fully specified by their mean and covariance functions. Second, it is necessary to specify a function to determine which combination of stimuli to propose next, typically refered to as the acquistion function [25]. In the present work, the expected improvement acquistion function was employed. Informally, this choice of acquistion can be seen as trying to maximize the expected improvement (in this case the increase in BOLD difference between the target regions) over the current best. When combined with a Gaussian process prior, such an acquisition function has a closed form solution; allowing the algorithm to rapidly propose new stimuli in real-time.

#### 4.2.1. Algorithmic details

The Bayesian optimisation procedure, summarized in Algorithm 1, is computationally and mathematically simple. An initial burn-in phase of five randomly selected stimuli is employed. Thereafter, at each iteration a new combination of audio-visual stimuli is proposed by maximising the expected improvement acquisition function. The proposed stimuli are subsequently presented to the subject and the BOLD difference between the target regions is provided as feedback to the algorithm. This feedback is used to update the posterior distribution of the unknown objective function, as described in [27]. Due to the properties of Gaussian processes, this update is computationally efficient and can be computed in closed form [25].

1) Run initial burn-in with 5 randomly selected stimuli
2) For $t = 6, .., n_{max}$:
   a) Select next stimulus combination by maximizing expected improvement:
   $$x_t = x_{next} = argmax_x\{EI(x)\}$$
   b) Propose stimulus $x_t$ to subject and measure BOLD activation in target ROIs
   c) Given BOLD activation, update posterior distribution as described in [27]

**Algorithm 1:** Description of the proposed Bayesian Optimisation algorithm

See Video 2 for an example how the algorithm is exploring the experiment parameter space over time and is learning the relationship between target brain state (BOLD difference between the ROIs) and stimuli combinations. As can be seen, from iteration 12 on the algorithm seems to have obtained a global understanding of the experiment parameters space and keeps sampling the predicted optimum over multiple iterations, hence trying to maximize the expected improvement as described above.

### 4.2.2. Gaussian Process Covariance Function

The choice of covariance function when employing Gaussian processes is fundamental [25]. In this work a squared exponential kernel was employed:

$$k(x, y) = \sigma^2 \exp\left\{-\frac{(x-y)^2}{2\, l^2}\right\}$$

where $x, y \in \mathbb{R}^2$ correspond to the choice of audio-visual stimuli. The hyper-parameters $\sigma, l \in \mathbb{R}$ each determine the variance and length scale of the covariance kernel respectively and must be carefully selected. In addition to this covariance function, it is also assumed that observations are corrupted by white noise. This is characterized by constant variance, $\sigma^2_{noise}$.

The choice of these three parameters is fundamental to the success of Bayesian optimisation methods. While it is possible to tune these hyper-parameters in real-time, in this work the parameters where selected prior to running the experiments. Formally, the data from the first study involving the SPSA algorithm was employed to tune these parameters using Type-2 maximum likelihood [27]. This choice of hyper-parameters was then fixed for all subjects' scans.

### 4.2.3. Acquisition Function

Similar to the choice of hyper-parameters, the choice of acquisition function is paramount to the success of Bayesian optimisation algorithms. In this work, the expected improvement acquisition function was employed. This choice was motivated by recent empirical [28] and theoretical [29] results. An additional attractive aspect of the expected improvement acquisition function is that it has a closed form under a Gaussian process prior. This implies that new stimuli combinations can be rapidly selected in real-time.

We define $m(x)$ as the predictive mean for a point $x \in \mathbb{R}^2$ and $var(x)$ as the predictive variance (i.e., $m(x)$ represents the mean expected value of BOLD activation for given stimuli $x$, and similarly for $var(x)$). The expected improvement is defined as [25]:

$$EI(x) = (m(x) - x_{max})\Phi(z) + var(x)\phi(z)$$

where $\Phi$ and $\phi$ are defined as the cumulative and probability density functions for a standard normal distribution respectively and $x_{max}$ is the maximum activation observed. Finally, z is defined as:

$$z = \frac{m(x) - x_{max}}{var(x)}$$

At every iteration, the next combination of stimuli to be observed is selected by maximizing the expected improvement:

$$x_{next} = argmax_x\{EI(x)\}$$

### 4.2.4. Objective function

After the presentation of a single audio-visual stimulus, we calculate the difference in BOLD activation between the two target brain regions. For this purpose, we run separate GLMs on the previous 10 time points of the cleaned time courses extracted from the occipital cortex and temporal cortex mask. Each GLM consisted of an intercept term and one stimulus regressor, which was modelled by convolving a boxcar kernel with a canonical double-gamma HRF. We simply took the difference between the resulting regression coefficients and entered them into the Bayesian optimisation algorithm.

### 4.2.5. Stopping criteria

Bayesian optimisation algorithms are much more sophisticated than the previously used stochastic approximation algorithms (i.e., SPSA). Such algorithms effectively balance a trade-off between exploration and exploitation, which stochastic methods ignore. As a result, it is challenging (and less important) to define a convergence criterion for Bayesian optimisation methods. Consequently, each run in this study was terminated after 19 observations had been sampled. This corresponded to 5.27% of the experimental parameter space or 190 TRs (6.3 minutes). Even during this reduced time, the Bayesian Optimisation algorithms where able to effectively learn, which stimuli combination maximized BOLD activation, and map out the whole experimental parameter space.

### 4.3. Post-hoc whole-brain fMRI analysis

We performed post-hoc fMRI analyses to illustrate that expected patterns of brain activity corresponded to maxima and minima in the experimental parameter space. To do this, we used standard FSL [14] FEAT GLM to determine the effect of the optimal stimuli combinations vs. least optimal stimuli combinations on individual's whole-brain BOLD activation. For this purpose we re-ran our Bayesian optimisation with all available observations per subject (i.e., concatenating all runs). The resulting model was then used to predict the entire experiment parameter space (361 possible states). We identified the optimum of the estimated experiment parameter space as the coordinate with the maximum predicted value by our model. In addition, we identified the coordinate with the minimum predicted value, representing the least optimal stimuli combination for evoking the pre-defined target brain state. Based on these results, we created three regressors that informed the GLM. For the first regressor, we determined all available observations for the respective subject that were in close vicinity to the optimum. For the second regressor we identified all available observations that were in vicinity to the least optimal stimuli combination. Note, that as the real-time stimuli selection using the Bayesian optimisation method is aiming to maximise the expected improvement, the least optimal stimuli combinations are undersampled compared to the most optimal stimuli combinations. All remaining observations were input for the third regressor. An observation was modelled as 10 second long boxcar and subsequently convolved with a double-gamma HRF. In addition, each regressors' first temporal derivative was included in the GLM. This first-level analysis was carried out for each run separately. The resulting parameter estimates of each run were then entered into a higher-level (fixed-effect) cluster-corrected FEAT analysis to summarize the results per individual with respect to the two contrast: "most optimal stimuli combinations > least optimal stimuli combinations" and "least optimal stimuli combinations > most optimal stimuli combinations". All final subject-level images were thresholded using a cluster correction threshold of nominal $z > 2.3$ and a nominal cluster significance threshold of $p = 0.05$. Group-level images were visualized on an average surface brain using MRIcroGL (http://www.mccauslandcenter.sc.edu/mricrogl/). Given that we are contrasting observations classified based on a prior statistical comparison (i.e., the Bayesian optimisation results) statistical results should be considered biased. Instead they are presented here for illustrative purposes so the approximate spatial distribution of activation and individual variability can be observed.

### 4.4. Results

*Defining the parameter space and finding the optimum*

First, we present results summarizing all four runs for each participant by re-estimating the predicted experiment parameter space post-hoc based on all available observations per subject (19 x 4 observations for four subjects, 19 observations for sub_04). The results of this analysis are depicted in Figure 4. The predicted optimum was defined as the coordinate with the maximum predicted value (see red dashed lines in Figure 4a). As can be clearly seen, the Bayesian optimisation method provides an optimum close to the hypothesized optimal combination of audio-visual stimuli for each subject: with the most optimal audio-visual stimulus combinations in the centre of the grid (maximum predicted value) and the least optimal stimuli combinations in each of the grid's corners (minimum predicted value). Across all subjects, the mean ± SD Euclidean distance between the empirical optimum and the hypothesized optimum at [10 10] was 1.48 ± 0.87.

In addition, a post-hoc whole-brain fMRI analysis was performed. The orange dots in Figure 4b were entered into the GLM as observations for the regressor modelling the most optimal stimulus combinations while the dark blue dots entered the GLM as least optimal stimulus combinations. The cluster corrected results of the higher-level (summarized over all runs) analysis are shown in Figure 4c for each participant. The contrast "most optimal stimuli combinations > least optimal stimuli combinations" is shown in yellow while the contrast "least optimal stimuli combinations > most optimal stimuli combinations" is rendered in dark blue.

As can be seen, our target brain regions of bilateral lateral occipital cortex and bilateral superior temporal cortex were well suited for capturing the desired effects. This is of no surprise as our ROIs were based on a previous study [13] and strongly activate for complex visual or auditory stimuli, respectively. This unconstrained whole-brain analysis though, provides us with insight about other brain regions that map similarly onto the experiment parameter space as the target brain regions. For example, from this we observe that early visual cortices (V1/V2), which were not included in the target occipital mask, are behaving in a comparable fashion in response to the presented stimuli as the higher visual regions in this case which were included in the target mask.

Most importantly, we observe some inter-individual differences. While the Bayesian optimisation algorithm converged to a very confined representation of the parameter space with a clear peak in the centre of the grid for sub_01, sub_02 and sub_05, the estimated parameter space for sub_03 appears much more distributed and less clear (see Figure 4a). Although, the algorithm correctly found the optimum to be located at [10 10], the available observations of this subject around the optimum are actually more widespread compared to other subjects (see Figure 4b). This indicates that for that subject's individual runs, the algorithm was less certain about the location of the optimum and so sampled more varied points in the parameter space. This is also reflected in the whole-brain fMRI result, which clearly shows weaker contrast activations than for the other subjects. Interestingly, although the algorithm was clearly exposed to very noisy and/or weaker differences in BOLD activation, it correctly converged to the optimum when taking all observations into account from this subject.

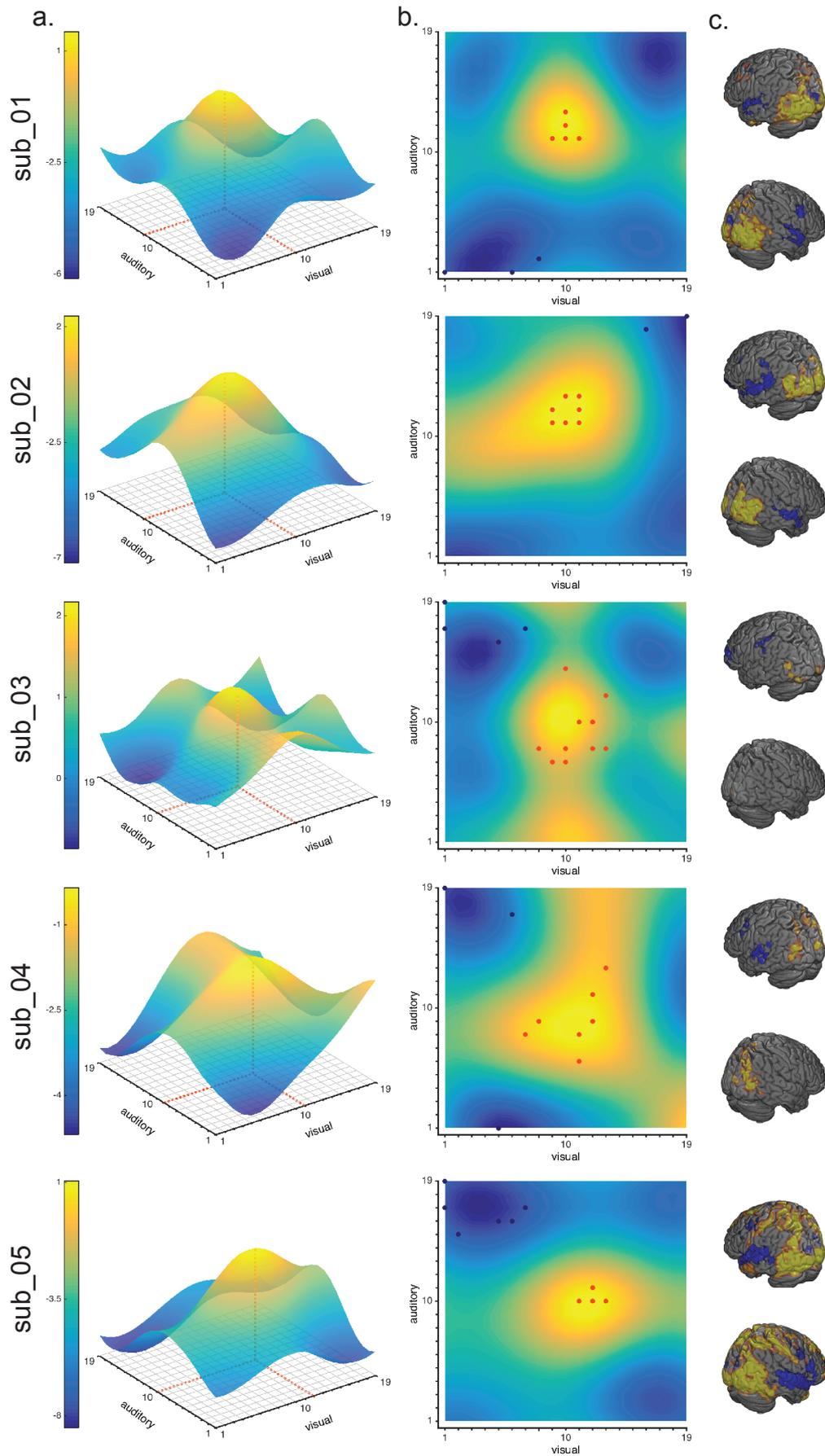

**Figure 4:** Individual estimated parameter space (a. and b.) and post-hoc fMRI pattern of activation for each subject (c.). The data summarizes results from all four runs except for sub_04 who only completed a single run due to MRI technical failure during scanning.

*Model performance in real-time*

Second, we were interested in assessing the online (rather than aggregate) performance of the Bayesian optimisation approach for each run separately to judge how new information updates the model. Therefore, in addition to the Euclidean distance, we also analysed the uncertainty of the algorithm, captured in the standard deviation of the algorithm's predictions. We derived both measures at each iteration of each run, by updating our model with every new observation made, i.e., incrementally re-simulating the online scenario. The updated model was then used to obtain a prediction of the activation response for every possible pair of combinations (361 pairs in total) for each iteration. The first five observations of each run were used as a burn-in for a first estimate of the model. Figure 5a depicts the mean standard deviation for each iteration of each run for all subjects (one colour per subject). As expected, uncertainty of our predictions decreased with an increasing number of observations, thus demonstrating that the algorithm is learning in real-time. Inter- and intra-individual differences seem negligible. Further, we found that the level of uncertainty stabilizes towards the end of the run and little to no additional decrease in uncertainty is evident with more observations.

For each subject, we identified the optimum as the coordinate that maximized the predicted value. We were then able to calculate the Euclidean distance of this coordinate from the hypothesized optimum in the centre of the grid. This analysis was performed at each iteration using the updated model. We then averaged the Euclidean distance across all runs for each subject (Figure 5b). Given that there was considerable variance across the runs for each subject, we analysed each run separately (averaging across all subjects) (Figure 5c). Interestingly, the first and second runs outperform the last two runs with respect to a smaller mean Euclidean distance from the optimum as well as smaller standard deviation towards the end of these runs. This finding may relate to habituation to the audio-visual stimuli and/or enhanced tiredness/boredom in part of the subjects in the last two runs.

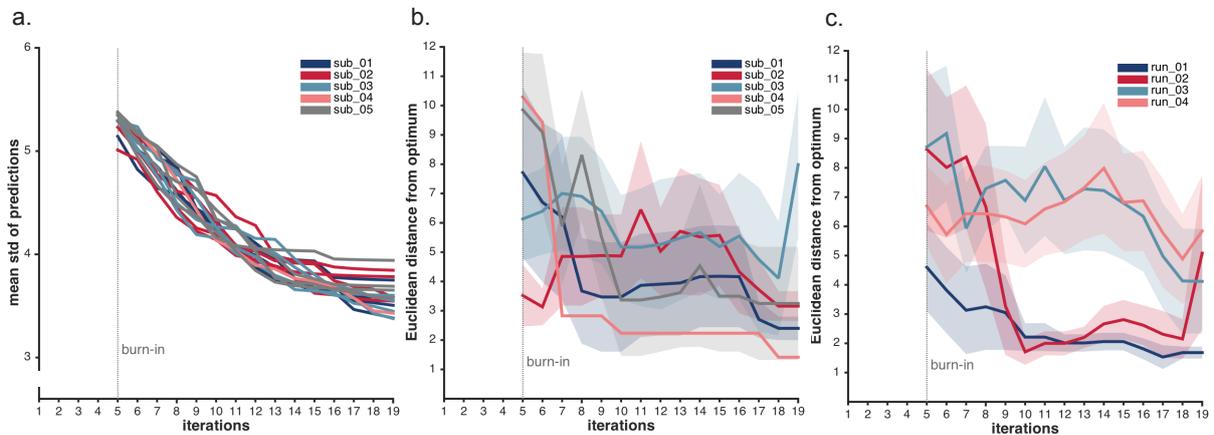

**Figure 5:** Summary measures of Bayesian model performance in real-time at each iteration. **a.** Mean standard deviation (std) of all predicted values for each run of every subject (one colour per subject). **b.** Mean Euclidean distance from hypothesized optimum across runs for each subject. The shaded areas represent the standard error. **c.** Mean Euclidean distance from hypothesized optimum across subjects for each run. The shaded areas represent the standard error.

*Assessing how many observations and runs are needed*

In a third analysis, we assessed how the Bayesian model estimate evolves over time, when not treating each run independently, but rather updating the model from the first run with each new observation made in succeeding runs from the same subject. In this way, the model estimated following the first run was used as the prior model for the second run, and so on. The results of this analysis are shown in Figure 6 for four subjects, excluding sub_04 as only one run was available. Note, that at the final update (after observation 75), the results are the same as those depicted in Figure 4a. It can be observed that there is a rapid learning gradient, i.e. decreased uncertainty, over the first run while it levels off for the last two runs (see Figure 6a). Further, we found that for all subjects the minimum Euclidean distance was obtained at the end of the first run or over the course of the second run (see Figure 6b). Note that a large Euclidean distance in the beginning of the first run is not disadvantageous as the algorithm is still exploring the parameter space at this stage (as the case for sub_02). While for three subjects, no further optimisation was achieved when adding observations made in the second, third and fourth run, we even find an adverse effect for one subject (sub_03) in the last two runs. These results also neatly correspond to our findings with respect to sub_03 mentioned above (Figure 4a-c). The last two runs for this subject seemed corrupted by very noisy measurements that lead to sudden changes in the algorithm's experimental parameter estimates.

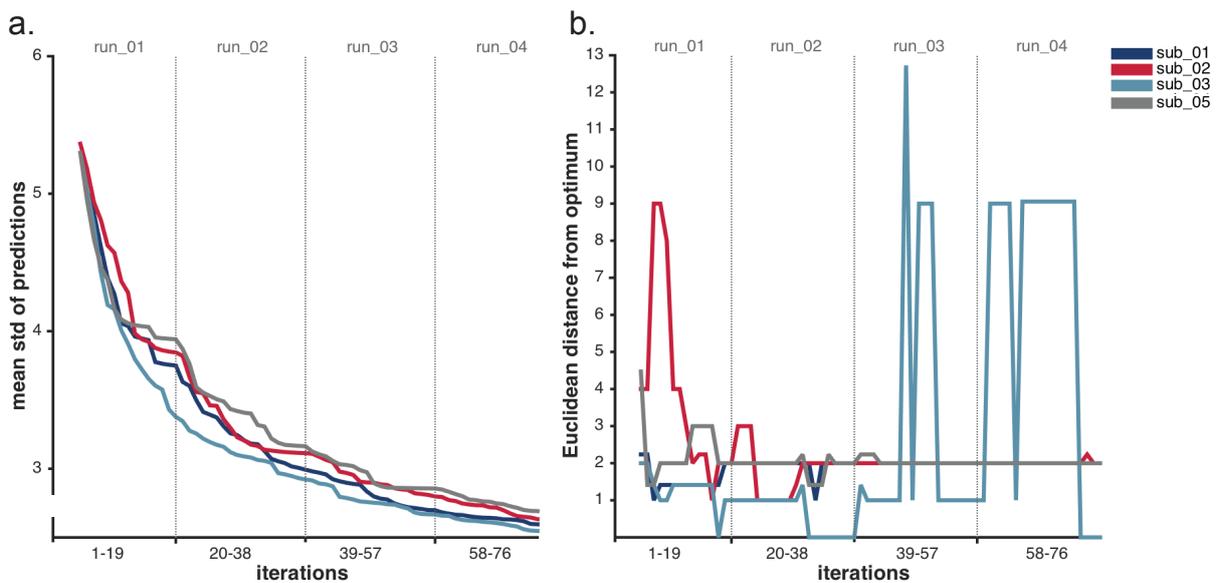

**Figure 6:** Summary measures of Bayesian model performance at each iteration when each new run of a subject is treated as new observation. Results are shown for four subjects as only one run was available for sub_04. **a.** Mean standard deviation of predicted values (std). **b.** Euclidean distance of predicted optimum from hypothesized optimum.

Based on these results, we concluded that there was relatively little or no gain (in terms of reduction of uncertainty or distance from theoretical optimum) from adding a second, third or fourth run to our experiment. However, this effect is based on the specific experimental conditions considered here, and we predict the effect will vary depending on the actual experimental conditions and target regions chosen.

This finding is also emphasized when estimating the experiment parameter space based only on the first runs of each subject. For this purpose, the Bayesian model was not re-estimated for each subject independently, but the model was updated with treating succeeding runs of the other subjects as new observations. Using this approach, the model converged to a predicted optimum at coordinate [9 10] (see Figure 7a); corresponding to a Euclidean distance of 1. This finding was identical to the minimum Euclidean distance when using all available runs of all subjects (Figure 7b). In this case, the predicted optimum was located at the coordinate [10 11]. Also when quantitatively assessing the similarity between the estimated parameter space based on only the first runs vs. all runs of all subjects, we find a very high spatial correlation ($r = .92$). Furthermore, for both estimated parameter spaces separately, we computed the spatial correlation between the final estimation (at iteration 95 or 323, respectively) and each previous iteration (from observation 5 on). The results of these analyses are depicted on the right side of Figure 7a or Figure 7b, respectively. We find that, for both models, the similarity to the final parameter space estimates is consistently increasing when including the first three subjects with only marginal improvements for sub_04 and sub_05.

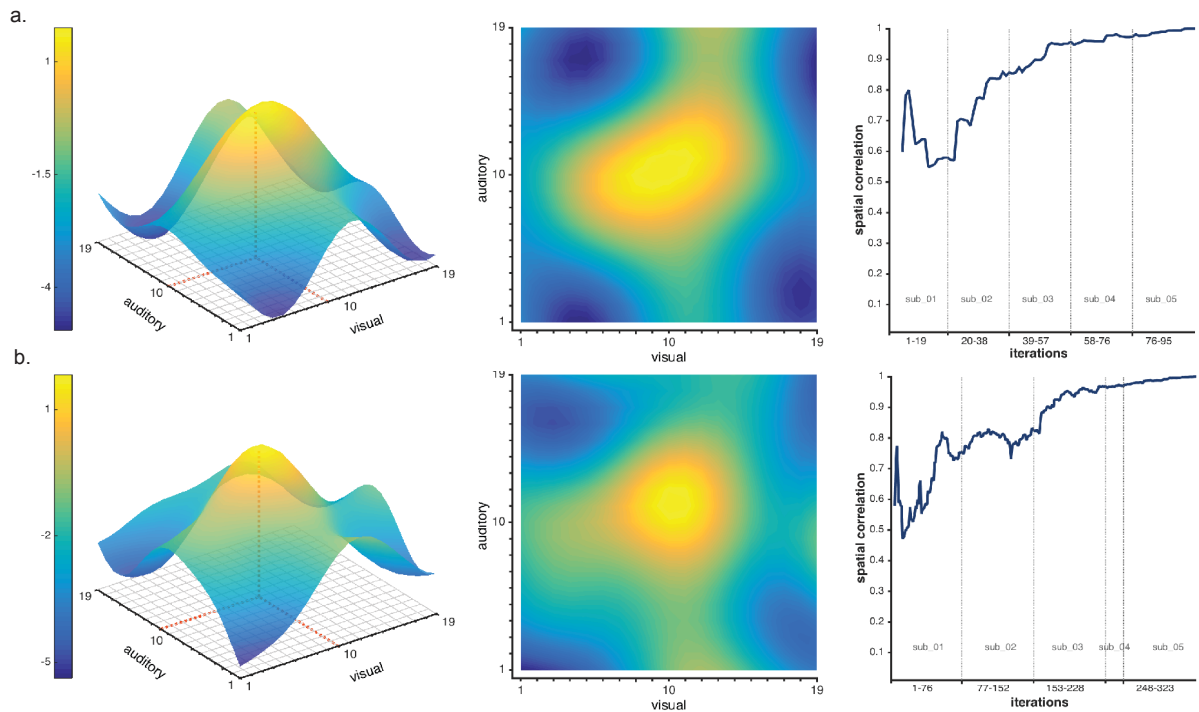

**Figure 7:** Parameter space estimates based on only the first runs of all subjects (**a.**) and all runs of all subjects (**b.**). The spatial correlations between the final parameter space estimates at iteration 95 or 323, respectively, and each previous observation (from iteration 5 on) is depicted on the right side.

In order to assess the statistical significance of our group-level results obtained when using only the first run of each subject we performed non-parametric permutation testing. For this purpose, we created a null-distribution by simulation the Euclidean distance of the predicted optimum when shuffling our empirically obtained objective function values (difference between the regression coefficients at each observation). The final model after 95 iterations (19 x 5 subject observations) was then used to predict the entire experiment parameter space. Finally, the Euclidean distance of the predicted optimum from the hypothesized optimum was calculated. This simulation was run for 10,000 iterations in order to obtain a distribution of test values expected under the null-hypothesis. When testing our empirically obtained Euclidean distance of 1 against this distribution, we obtained a *z*-value of -3.51 ($p = 0.00015$, one-tailed).

## 5. Discussion

The present work demonstrates the feasibility of our proposed *automatic neuroscientist* approach, turning on its head how a typical fMRI experiment is carried out. The results from the first study show that the approach can rapidly and accurately adjust the experimental conditions in real-time in order to maximize similarity with a target pattern of brain activity. We achieved convergence at the hypothesized optimal audio-visual stimulus combination in 11 out of 14 runs in less than 10 minutes. In comparison, if we had sampled each of the 100 possible states, the scanning time would have been a minimum of half an hour per subject (100 states x 20 seconds). However, as mentioned above, the algorithm employed in the first study only aims for convergence but does not provide the experimenter with a description of the underlying experiment parameter space. In the second study, by employing a Bayesian optimisation approach we demonstrated that we could rapidly obtain an accurate estimation of the whole experimental parameter space. We showed that using the first run from the five subjects provided us with a statistically significant and accurate mapping of the hypothesized optimum. Each run only lasted 6.3 minutes, while it would have taken more than two hours scanning time for each individual to exhaustively test all combinations in the parameter space (361 states x 20 seconds). This approach, therefore, provides us with an efficient and distinct way of using fMRI to explore the relationship between tasks and the brain, with a number of different, potential applications (detailed below).

In the following paragraphs we describe four different potential scenarios where the *Automatic Neuroscientist* would be useful:

**Scenario 1**: Opening new avenues in cognitive neuroscience

While the current study demonstrates the feasibility of our approach, using perceptual stimuli, we eventually aim to apply the *Automatic Neuroscientist* to higher-level cognitive tasks. Our method provides a novel tool to address the other side of the many tasks to many regions mapping by understanding the relationship on how many different cognitive tasks can activate the same brain system. This perspective on investigating how the brain is associated with cognition is overlooked in the vast majority of imaging studies. To date, this question can only be partially addressed by using post-hoc meta-analyses combining many different studies involving different tasks [11]. Limitations of meta-analyses are evident and range from differences in hardware (e.g. scanner strength) as well as pre-processing and analysis pipelines, to differences in experimental design, subject sample composition [30] and a coarse cognitive ontology in the databases (including the coarse coding of the experiments and the cognitive processes involved [11]). Most importantly, meta-analyses are based on published coordinates of group level results; hence inter-subject variability is considered as an effect to overcome. Our approach, however, strives to capture individual differences not in terms of regional activation differences for a given task, but in terms of how a set of tasks/stimuli relate to the region.

**Scenario 2**: Optimize stimuli in fMRI piloting

In contrast to the cognitive theoretical applications, above, our approach also has simple, practical potential benefits, when designing fMRI experiments. Typically, designing an experiment involves, at best, heuristics about many task and stimulus parameters (e.g., inter-stimulus-intervals (ISI), length of baselines, type of stimuli used, etc.). These are then assessed in a few pilot scans before embarking on a larger study involving many participants or comparing across individuals or groups. Using the approach detailed here, these pilot scans can be conducted in a more efficient and principled way, to find the set of e.g., stimuli/task parameters that maximally evoke the target brain response, before acquiring larger datasets, potentially dramatically improving the signal-to-noise ratio of resulting experiments.

**Scenario 3**: Tailoring clinical rehabilitation therapy to the patient

*Neurofeedback.* Recent technical advances have made real-time fMRI a tool for performing neurofeedback in a clinical context. Numerous promising studies in clinical population show symptomatic improvements in patients suffering from Parkinson's diseases, chronic pain, chronic tinnitus as well as schizophrenia, depression, post-traumatic stress disorder and addiction [31]. The main benefit may involve the long-term restoration of healthy brain activity, without the side effects commonly associated with chronic medication. The benefits of using experimenter-defined explicit strategies or subject-generated implicit strategies for self-regulation is an ongoing topic of debate [32]. However, our proposed framework could be employed to search efficiently and systematically through a variety of different explicit mental strategies. This would allow us to pinpoint the individual mental strategy found to reliably maximize brain activity in a given target brain region. This verbalization might drastically limit scanning time per individual and facilitate training sessions at home.

*Rehabilitative cognitive or behavioural therapy.* Of maybe more direct clinical relevance (given the high cost of MRI scanning), our approach could be useful in tailoring cognitive or behavioural rehabilitation to a specific patient's needs, for example, following stroke or traumatic brain injury. For instance, altered connectivity between default mode network and fronto-parietal networks during effortful cognitive tasks has been related to cognitive impairment [33]. Therefore, some type of computerized behavioural intervention can be optimized (i.e., parameters such as difficulty, ISI, task type) and tailored to individual patients, to develop a behavioural training regime designed to normalise dysfunctional connectivity. This tailored computer training could then be extensively delivered to the patient outside the scanner (e.g., at home), and potentially, improve cognitive function.

*Brain stimulation.* Brain stimulation has evolved into a rapidly advancing field of scientific research. Studies involving non-invasive electrical stimulation, such as transcranial direct current stimulation (tDCS) and transcranial alternating current stimulation (tACS), have reported promising modulation on cognitive and behavioural performance [34], [35]. Simultaneous fMRI is increasingly used to understand the functional brain networks affected by the intervention [36]–[38]. In most studies only a small range of possible stimulation parameters is varied and investigated. However, the optimal stimulation paradigm may vary across individual patients based on variability in the specific pathology of the disease as well as more general individual variability. We propose to tackle this issue by combining our approach with short periods of presentation of brain stimulation in the scanner and systematically explore a high-dimensional space of different stimulation parameters (including the stimulation site, intensity and duration of stimulation for tDCS, amplitude and phase for tACS) aimed at maximizing "healthy" patterns of brain activity for a specific patient. Similarly, adaptive deep brain stimulation (DBS) in the subthalamic nucleus (using local field potentials recorded directly from the stimulation electrodes) has been shown to be more efficient and efficacious that constant DBS [39]. Significant behavioural improvement was achieved with a reduction in stimulation time; hence highlighting the potential of personalized stimulation therapy.

**Scenario 4**: Assessing an individual's preferences
Another potential application of our proposed framework is to assess subjects' preferences for specific types of stimuli in an efficient way, and to design stimuli that maximally activate brain regions known to be involved in, e.g., reward. For example, stimulus features (such as a picture or sound) could be systematically explored and optimized towards a desired response in the brain when designing an advertisement; this could be done on different target groups, hence, facilitating group specific advertisement.

In order to achieve these potential applications, there are a number of areas for ongoing and future work. First, the efficiency at estimating the experiment parameter space will be a function of the complexity of the space and the specific paradigm used. We acknowledge that in some cases, the time in which the Bayesian optimisation algorithm will be able to accurately map out the experiment parameter space will vary as a function of the signal-to-noise ratio of the desired target brain state measure. For instance, for more subtle distinctions between relatively similar cognitive tasks, the signal-to-noise ratio is expected to drop and hence, the algorithm is expected to take longer to accurately estimate the parameter space. One avenue for future work, therefore, is developing online stopping criteria, which automatically ends the current run as soon as the uncertainty of the algorithm over the parameter space is sufficiently small. In a similar vein, using simulations and offline fMRI data, Feng and colleagues [40] recently demonstrated the potential for dynamically stopping stimulus presentation when sufficient statistical evidence is collected to determine activation in a given brain region.

Second, the *automatic neuroscientist* can be extended to any target brain state. While, in the present work, we have demonstrated a target brain state that is simply based on BOLD differences between two brain regions, any desirable brain state could be defined. For instance, the stimuli could be optimized to maximize a specific functional connectivity network configuration. We have recently presented a real-time extension of the Smooth Incremental Graphical Lasso Estimation (SINGLE, [41]), which captures dynamic changes in functional connectivity in real-time [42]. We are currently exploring combining the real-time SINGLE method with the Bayesian optimisation approach presented here, allowing for a richer description of a desired target brain state.

To our knowledge only one study has applied real-time fMRI to determine what stimuli are presented. In this study, Cusack et al. [43] employed online multivariate pattern analysis to converge to a subset of images that evoke a similar BOLD pattern than a reference image. This subset of images was then referred to as the references image's 'neural neighbourhood'. In another recent study, Feng et al. [40] address the possibility to dynamically adjust task difficulty levels in order to determine the minimum task difficulty level that will activate a given ROI but only simulations have been carried out. However, these studies do not provide a generalizable framework that is applicable to numerous different research questions.

With the work we present here, we aim to stimulate the field to turn fMRI on its head and explore the wide range of novel applications involving closed-loop real-time fMRI. We envision that the type of approach explained here, will be added to the standard toolkit of modern functional imaging.

# References

[1] K. D. Davis, "The neural circuitry of pain as explored with functional MRI," *Neurol. Res.*, vol. 22, no. 3, pp. 313–317, Apr. 2000.

[2] R.-D. Treede, D. R. Kenshalo, R. H. Gracely, and A. K. P. Jones, "The cortical representation of pain," *PAIN*, vol. 79, no. 2–3, pp. 105–111, Feb. 1999.

[3] T. D. Wager, L. Y. Atlas, M. A. Lindquist, M. Roy, C.-W. Woo, and E. Kross, "An fMRI-Based Neurologic Signature of Physical Pain," *N. Engl. J. Med.*, vol. 368, no. 15, pp. 1388–1397, Apr. 2013.

[4] W. W. Seeley, V. Menon, A. F. Schatzberg, J. Keller, G. H. Glover, H. Kenna, A. L. Reiss, and M. D. Greicius, "Dissociable Intrinsic Connectivity Networks for Salience Processing and Executive Control," *J. Neurosci.*, vol. 27, no. 9, pp. 2349–2356, Feb. 2007.

[5] L. Q. Uddin, "Salience processing and insular cortical function and dysfunction," *Nat. Rev. Neurosci.*, vol. 16, no. 1, pp. 55–61, Jan. 2015.

[6] G. Hein and R. T. Knight, "Superior temporal sulcus--It's my area: or is it?," *J. Cogn. Neurosci.*, vol. 20, no. 12, pp. 2125–2136, Dec. 2008.

[7] A. Amedi, K. von Kriegstein, N. M. van Atteveldt, M. S. Beauchamp, and M. J. Naumer, "Functional imaging of human crossmodal identification and object recognition," *Exp. Brain Res.*, vol. 166, no. 3–4, pp. 559–571, Oct. 2005.

[8] A. Puce and D. Perrett, "Electrophysiology and brain imaging of biological motion.," *Philos. Trans. R. Soc. B Biol. Sci.*, vol. 358, no. 1431, pp. 435–445, Mar. 2003.

[9] C. J. Price, "The anatomy of language: contributions from functional neuroimaging," *J. Anat.*, vol. 197, no. Pt 3, pp. 335–359, Oct. 2000.

[10] J. V. Haxby, L. G. Ungerleider, V. P. Clark, J. L. Schouten, E. A. Hoffman, and A. Martin, "The effect of face inversion on activity in human neural systems for face and object perception," *Neuron*, vol. 22, no. 1, pp. 189–199, Jan. 1999.

[11] R. A. Poldrack, "Can cognitive processes be inferred from neuroimaging data?," *Trends Cogn. Sci.*, vol. 10, no. 2, pp. 59–63, Feb. 2006.

[12] C. F. Ferris, P. Kulkarni, J. M. Sullivan, J. A. Harder, T. L. Messenger, and M. Febo, "Pup Suckling Is More Rewarding Than Cocaine: Evidence from Functional Magnetic Resonance Imaging and Three-Dimensional Computational Analysis," *J. Neurosci.*, vol. 25, no. 1, pp. 149–156, Jan. 2005.

[13] R. M. Braga, L. R. Wilson, D. J. Sharp, R. J. S. Wise, and R. Leech, "Separable networks for top-down attention to auditory non-spatial and visuospatial modalities," *NeuroImage*, vol. 74, pp. 77–86, Jul. 2013.

[14] M. Jenkinson, C. F. Beckmann, T. E. J. Behrens, M. W. Woolrich, and S. M. Smith, "FSL," *NeuroImage*, vol. 62, no. 2, pp. 782–790, Aug. 2012.

[15] S. M. Smith, "Fast robust automated brain extraction," *Hum. Brain Mapp.*, vol. 17, no. 3, pp. 143–155, Nov. 2002.

[16] M. Jenkinson and S. Smith, "A global optimisation method for robust affine registration of brain images," *Med. Image Anal.*, vol. 5, no. 2, pp. 143–156, Jun. 2001.

[17] M. Jenkinson, P. Bannister, M. Brady, and S. Smith, "Improved optimization for the robust and accurate linear registration and motion correction of brain images," *NeuroImage*, vol. 17, no. 2, pp. 825–841, Oct. 2002.

[18] Y. Koush, M. Zvyagintsev, M. Dyck, K. A. Mathiak, and K. Mathiak, "Signal quality and Bayesian signal processing in neurofeedback based on real-time fMRI," *NeuroImage*, vol. 59, no. 1, pp. 478–489, Jan. 2012.

[19] D. H. Brainard, "The Psychophysics Toolbox," *Spat. Vis.*, vol. 10, no. 4, pp. 433–436, 1997.

[20] D. G. Pelli, "The VideoToolbox software for visual psychophysics: transforming numbers into movies," *Spat. Vis.*, vol. 10, no. 4, pp. 437–442, 1997.

[21] J. B. Millar, J. P. Vonwiller, J. M. Harrington, and P. J. Dermody, "The Australian National Database of Spoken Language," in *, 1994 IEEE International Conference on Acoustics, Speech, and Signal Processing, 1994. ICASSP-94*, 1994, vol. i, pp. I/97–I100 vol.1.
[22] P. Boersma, "Praat, a system for doing phonetics by computer.," *Glot Int.*, vol. 5, no. 9/10, pp. 341–345, 2001.
[23] J. C. Spall, "An Overview of the Simultaneous Perturbation Method for Efficient Optimization," *John Hopkins APL Tech. Dig.*, vol. 19, no. 4, pp. 482–492, 1998.
[24] L. Gerencsér, D. H. Stacy, and Z. S. Vágó, "Optimization over discrete sets via SPSA," in *Proceedings of the 38th IEEE Conference on Decision and Control*, Phoenix, Arizona, 1999, pp. 1791–1794.
[25] E. Brochu, V. M. Cora, and N. de Freitas, "A Tutorial on Bayesian Optimization of Expensive Cost Functions, with Application to Active User Modeling and Hierarchical Reinforcement Learning," *ArXiv10122599 Cs*, Dec. 2010.
[26] J. Snoek, H. Larochelle, and R. P. Adams, "Practical Bayesian Optimization of Machine Learning Algorithms," in *Advances in Neural Information Processing Systems 25 (NIPS 2012)*, F. Pereira, C. J. C. Burges, L. Bottou, and K. Q. Weinberger, Eds. Curran Associates, Inc., 2012, pp. 2951–2959.
[27] C. E. Rasmussen and C. K. I. Williams, *Gaussian Processes for Machine Learning*. Cambridge, Mass: MIT Press, 2006.
[28] M. Osborne, "Bayesian Gaussian Processes for Sequential Prediction, Optimisation and Quadrature," Dissertation Thesis, University of Oxford, 2010.
[29] A. D. Bull, "Convergence Rates of Efficient Global Optimization Algorithms," *J Mach Learn Res*, vol. 12, pp. 2879–2904, Nov. 2011.
[30] S. G. Costafreda, "Pooling fMRI Data: Meta-Analysis, Mega-Analysis and Multi-Center Studies," *Front. Neuroinformatics*, vol. 3, Sep. 2009.
[31] S. Ruiz, K. Buyukturkoglu, M. Rana, N. Birbaumer, and R. Sitaram, "Real-time fMRI brain computer interfaces: Self-regulation of single brain regions to networks," *Biol. Psychol.*, vol. 95, pp. 4–20, Jan. 2014.
[32] J. Sulzer, S. Haller, F. Scharnowski, N. Weiskopf, N. Birbaumer, M. L. Blefari, A. B. Bruehl, L. G. Cohen, R. C. deCharms, R. Gassert, R. Goebel, U. Herwig, S. LaConte, D. Linden, A. Luft, E. Seifritz, and R. Sitaram, "Real-time fMRI neurofeedback: Progress and challenges," *NeuroImage*, vol. 76, pp. 386–399, Aug. 2013.
[33] S. R. Jilka, G. Scott, T. Ham, A. Pickering, V. Bonnelle, R. M. Braga, R. Leech, and D. J. Sharp, "Damage to the Salience Network and Interactions with the Default Mode Network," *J. Neurosci.*, vol. 34, no. 33, pp. 10798–10807, Aug. 2014.
[34] M.-F. Kuo and M. A. Nitsche, "Effects of transcranial electrical stimulation on cognition," *Clin. EEG Neurosci.*, vol. 43, no. 3, pp. 192–199, Jul. 2012.
[35] M. Mondino, D. Bennabi, E. Poulet, F. Galvao, J. Brunelin, and E. Haffen, "Can transcranial direct current stimulation (tDCS) alleviate symptoms and improve cognition in psychiatric disorders?," *World J. Biol. Psychiatry Off. J. World Fed. Soc. Biol. Psychiatry*, vol. 15, no. 4, pp. 261–275, May 2014.
[36] A. Antal, R. Polania, C. Schmidt-Samoa, P. Dechent, and W. Paulus, "Transcranial direct current stimulation over the primary motor cortex during fMRI," *NeuroImage*, vol. 55, no. 2, pp. 590–596, Mar. 2011.
[37] C. Saiote, Z. Turi, W. Paulus, and A. Antal, "Combining functional magnetic resonance imaging with transcranial electrical stimulation," *Front. Hum. Neurosci.*, vol. 7, Aug. 2013.
[38] B. Sehm, A. Schäfer, J. Kipping, D. Margulies, V. Conde, M. Taubert, A. Villringer, and P. Ragert, "Dynamic modulation of intrinsic functional connectivity by transcranial

direct current stimulation," *J. Neurophysiol.*, vol. 108, no. 12, pp. 3253–3263, Dec. 2012.

[39] S. Little, A. Pogosyan, S. Neal, B. Zavala, L. Zrinzo, M. Hariz, T. Foltynie, P. Limousin, K. Ashkan, J. FitzGerald, A. L. Green, T. Z. Aziz, and P. Brown, "Adaptive deep brain stimulation in advanced Parkinson disease," *Ann. Neurol.*, vol. 74, no. 3, pp. 449–457, Sep. 2013.

[40] I. J. Feng, A. I. Jack, and C. Tatsuoka, "Dynamic adjustment of stimuli in real time functional magnetic resonance imaging," *PloS One*, vol. 10, no. 3, p. e0117942, 2015.

[41] R. P. Monti, P. Hellyer, D. Sharp, R. Leech, C. Anagnostopoulos, and G. Montana, "Estimating time-varying brain connectivity networks from functional MRI time series," *NeuroImage*, vol. 103, pp. 427–443, Dec. 2014.

[42] R. P. Monti, R. Lorenz, C. Anagnostopoulos, R. Leech, and G. Montana, "Measuring the functional connectome 'on-the-fly': towards a new control signal for fMRI-based brain-computer interfaces," *ArXiv150202309 Stat*, Feb. 2015.

[43] R. Cusack, M. Veldsman, L. Naci, D. J. Mitchell, and A. C. Linke, "Seeing different objects in different ways: Measuring ventral visual tuning to sensory and semantic features with dynamically adaptive imaging," *Hum. Brain Mapp.*, vol. 33, no. 2, pp. 387–397, 2012.